\documentclass[12pt]{iopart}
\newtheorem{thm}{Theorem}

\newtheorem{pro}{Proposition}

\usepackage{iopams}
\begin{document}

\title[On some classes of discrete polynomials and ordinary difference equations]{On some classes of discrete polynomials and ordinary difference equations}

\author{Andrei K Svinin}

\address{Institute for System
Dynamics and Control Theory, Siberian Branch of
Russian Academy of Sciences, Russia}
\ead{svinin@icc.ru}
\begin{abstract}
We introduce two classes of discrete polynomials and use them for constructing  ordinary difference equations admitting a Lax representation in terms of these polynomials.  We also construct lattice integrable hierarchies in their explicit form and show some examples.
\end{abstract}
\vspace{2pc}


\submitto{J. Phys. A: Math. Theor.}


\section{Introduction}

A discrete counterpart of an $N$th order autonomous ordinary differential equation is a recurrence relation of the form
\begin{equation}
T(i+N)=F(T(i), T(i+1),\ldots, T(i+N-1))
\label{ODE}
\end{equation}
for an unknown function $T(i)$ of a discrete variable $i\in\mathbb{Z}$. One calls (\ref{ODE}) an ordinary discrete (difference) equation of $N$th order. Equation (\ref{ODE}) yields a map $\mathbb{R}^N\rightarrow \mathbb{R}^N$ for real-valued initial data $\{y_j\equiv T(i_0+j) : j=1,\ldots, N-1\}$. 
By analogy with ordinary differential equations, the function $J=J(i)=J(T(i), T(i+1),\ldots, T(i+N-1))$ is called the first integral for difference equation (\ref{ODE}) if by virtue of this equation one has $J(i+1)=J(i)$. There are ordinary difference equations which have some special properties which yield a regular behavior of their solutions. One of such properties is complete or Liouville-Arnold integrability \cite{Bruschi}, \cite{Veselov}, that is, the existence of a sufficient number of functionally independent integrals which are in involution with respect to a Poisson bracket. Usually such integrable equations are grouped into some infinite classes of their own kind. Examples to mention are the sine-Gordon, modified KdV, potential KdV and Lyness equations \cite{Kamp}, \cite{Kamp1}, \cite{Tran1}, \cite{Tran2}. It is worth remarking that these equations can be obtained as special reductions of partial difference equations for an unknown function of two discrete variables. Studies carried out, for example, in \cite{Hone}, \cite{Tran3}
shows that proving  the Liouville-Arnold integrability for the map under consideration is a quite difficult and complicated task. 
Another characteristic  of discrete equations claiming to have an adjective `integrable' is a Lax pair representation. This means that the system of equations appear as a condition of compatibility of two linear equations with some spectral parameters. Let us remark that the above mentioned  difference equations in fact possess this property. It should be related to the  complete integrability, how most likely it can be considered just as an indicator of Liouville-Arnold integrability. Also possible integrability criteria which might be applied to discrete equations are the zero algebraic entropy \cite{Bellon}, \cite{Hietarinta}, \cite{Veselov1} and singularity confinement \cite{Grammaticos}.

The main purpose of our paper is to exhibit some classes of ordinary autonomous difference equations having a Lax pair representation. In the general case, we obtain multi-field difference systems which might be completely integrable. Also we present an approach which allows us to construct many integrable hierarchies of evolution differential-difference equations in their explicit form. This goal is achieved by solving recurrence equations on the coefficients of formal pseudo-difference operators in terms of which we write the Lax pairs. As a result we obtain some polynomials $T^k_s$ and $S^k_s$ which, in a sense, generalize elementary and complete symmetric polynomials \cite{Macdonald}, respectively. Therefore, we construct our  ordinary difference equations and hierarchies of evolution differential-difference equations in terms of these polynomials. Then we compare these studies with our approach given in \cite{Svinin2}, \cite{Svinin3}, \cite{Svinin4}, \cite{Svinin5} which relates many lattice integrable hierarchies to the Kadomtsev-Petviashvili (KP) hierarchy and the $n$th discrete KP hierarchy. It is worth remarking that this approach goes back to \cite{Magri}.  In this framework lattice integrable hierarchies appear as reductions of the $n$th discrete KP hierarchy on corresponding invariant submanifolds. It allows us to show  the meaning of obtained ordinary difference equations from this point of view. Namely, they turn out to be algebraic constraints compatible with corresponding integrable hierarchies which yield some invariant submanifolds in the solution space of the system under consideration. Looking ahead let us show the simplest example --- the Volterra lattice hierarchy of evolution equations\footnote{Here the symbol $\partial_s$ stands for derivative $\partial/\partial t_s$ with respect to the evolution parameter $t_s$.}
\[
\partial_sT(i)=(-1)^sT(i)\left\{S^s_s(i-s+2)-S^s_s(i-s)\right\},\;\;\;s\geq 1,
\]
whose right-hand sides are defined by the following discrete polynomials
\[
S^1_1(i)=T(i),\;\;\;
S^2_2(i)=T(i+1)\left\{T(i)+T(i+1)+T(i+2)\right\},\;\;\;
\]
\[
S^3_3(i)=T(i+2)\left\{T(i+1)\left\{T(i)+T(i+1)+T(i+2)+T(i+3)\right\} \right.
\]
\[
+T(i+2)\left\{T(i+1)+T(i+2)+T(i+3)\right\} 
\]
\[
\left. +T(i+3)\left\{T(i+2)+T(i+3)\right\}+T(i+4)T(i+3)\right\}
\]
and so on. This hierarchy corresponds to the invariant submanifold $\mathcal{M}_{1, 2, 1}$ of the discrete KP hierarchy \cite{Svinin1}. Our approach gives an infinite number of constraints
\[
\tilde{S}^k_{s+1}(i)T(i+k)=\tilde{S}^k_{s+1}(i+1)T(i+s),\;\;\;
s\geq k+1,\;\;\;
k\geq 0
\]
each of which represents an autonomous difference equation (\ref{ODE}) of the order $N=k+s$ with the right-hand side rationally depending on its arguments and  some number of parameters. In particular, in the case $k=0$, it is specified as the periodicity condition $T(i+s)=T(i)$. 

The rest of the paper is organized as follows. In section \ref{sect1}, in the framework of formal pseudo-difference operators, we construct two classes of above-mentioned quasi-homogeneous discrete polynomials of an infinite number of the fields $\{T^j=T^j(i) : j\geq 1\}$. All statements in this section we formulate in the language of corresponding multi-variate polynomials.  In section \ref{sect2}, we apply these polynomials for constructing ordinary difference equations which arise in this approach as a conditions of the consistency of two linear equations yielding the Lax pairs. We also construct integrable hierarchies of evolution differential-difference equations in their explicit form and provide the reader by well-known examples. In section \ref{sect3}, we set out our approach to integrable lattices from the point of view of the KP hierarchy and derive the same classes of ordinary difference equations in this framework. This enable us to explain the meaning of the difference equations obtained in section \ref{sect2} as some algebraic constraints compatible with the flows of corresponding integrable  hierarchies.


\section{Discrete polynomials $T^k_s$ and $S^k_s$}
\label{sect1}

The main goal of  this section is to describe some quasi-homogeneous polynomials associated with  formal pseudo-difference operators. This will enable us to construct ordinary difference equations admitting a Lax representation and integrable hierarchies of lattice evolution equations in their explicit form.


\subsection{Formal pseudo-difference operators and polynomials $T^k_s$ and $S^k_s$}

Given an infinite number of unknown functions $\{T^j=T^j(i) : j\geq 1\}$ of a discrete variable $i\in\mathbb{Z}$ and an arbitrary pair of co-prime integers  $h\geq 1$ and $n\geq 1$, we consider a pseudo-difference operator\footnote{Here $\Lambda$ stands for the shift operator acting as $\Lambda(f(i))=f(i+1)$.}
\[
\mathcal{T}_1=\Lambda^{-h}+\sum_{j\geq 1}z^{-j}T^j(i-(j-1)n-h)\Lambda^{-h-jn}
\]
and its powers\footnote{Strictly speaking, we have to indicate dependence of polynomials $T^k_s$ and $S^k_s$ on $n$ and $h$ writing, for example $T^{(n, h, k)}_s$ and $S^{(n, h, k)}_s$ but we believe that such notations are quite cumbersome and prefer to use simplified ones in the hope that this does not lead to a confusion.}
\begin{equation}
\mathcal{T}_s\equiv (\mathcal{T}_1)^s=\Lambda^{-sh}+\sum_{j\geq 1}z^{-j}T^j_{s}(i-(j-1)n-sh)\Lambda^{-sh-jn}
\label{PDO}
\end{equation}
for all $s\geq 1$. By definition, the coefficients $T^k_s$, for all $s\geq 1$  are some  discrete polynomials of infinite number of fields $T^j$. We indicate this as $T^k_s=T^k_s[T^1,\ldots, T^k]$.\footnote{We use the term `discrete polynomial'  by analogy with the notion of differential polynomial.}

By definition, $\mathcal{T}_{s_1+s_2}=\mathcal{T}_{s_1}\mathcal{T}_{s_2}=\mathcal{T}_{s_2}\mathcal{T}_{s_1}$. In terms of the coefficients of pseudo-difference operators (\ref{PDO}) this looks as a pair of two compatible identities
\begin{eqnarray}
T^k_{s_1+s_2}(i)&=&T^k_{s_1}(i+s_2h)+\sum_{j=1}^{k-1}T^j_{s_2}(i)T^{k-j}_{s_1}(i+s_2h+jn)+T^k_{s_2}(i) \nonumber
\\
&=&T^k_{s_2}(i+s_1h)+\sum_{j=1}^{k-1}T^j_{s_1}(i)T^{k-j}_{s_2}(i+s_1h+jn)+T^k_{s_1}(i). \nonumber
\end{eqnarray}
In particular, one sees that these polynomials must satisfy a compatible pair of recurrence relations
\begin{eqnarray}
\fl
T^k_s(i)&=&T^k_{s-1}(i+h)+\sum_{j=1}^{k-1}T^j(i)T^{k-j}_{s-1}(i+h+jn)+T^k(i) \label{res1} \\
\fl
        &=&T^k_{s-1}(i)+\sum_{j=1}^{k-1}T^j(i+(s-1)h+(k-j)n)T^{k-j}_{s-1}(i)+T^k(i+(s-1)h). \label{res2}
\end{eqnarray}
Taking for use one of them, one derives all polynomials $T^k_s$ starting from $T^k_1=T^k$.

Some remarks are in order. In practice, it is more convenient to work with multi-variate polynomials $Q\{y^r_j\}$ rather than with discrete ones identifying $y_j^r= T^r(i+j)$ for $r\geq 1$ and $j\in\mathbb{Z}$. In fact, it is not only the additional notation. We believe that multi-variate polynomials which we present below could be more fundamental objects than their discrete counterparts which are used throughout the paper for constructing discrete integrable systems having Lax pair representation. Therefore,  let us formulate in this section all propositions concerning discrete polynomials in the language of multi-variate polynomials.  

Let $k=[y^k_j]$ be a scaling dimension or weight. For monomials one puts
\[
\sum_{j=1}^rk_j=[y^{k_1}_{j_1}y^{k_2}_{j_2}\ldots y^{k_r}_{j_r}].
\]
One says that a polynomial $Q=Q\{y^r_j\}$ is quasi-homogeneous one of the degree $k$ if all monomials entering in this polynomial have the same weight $k$ or in other words that $Q\{\lambda^ry^r_j\}=\lambda^kQ\{y^r_j\}$ for any $\lambda\in\mathbb{R}$.  

Let us rewrite relations (\ref{res1}) and (\ref{res2}) as equivalent relations for multi-variate polynomials $T^k_s\{y^r_j\}$\footnote{The notations like $T^{k, \alpha}_s$ and $S^{k, \alpha}_s$ stand for $\alpha$-shifted polynomials, that is, $T^{k, \alpha}_s\equiv T^k_s\{y^r_{j+\alpha}\}$ and $S^{k, \alpha}_s\equiv S^k_s\{y^r_{j+\alpha}\}$, respectively.}
\begin{eqnarray}
T^k_s&=&T^{k,h}_{s-1}+\sum_{j=1}^{k-1}y^j_0T^{k-j, h+jn}_{s-1}+y^k_0 \label{res3} \\
&=&T^k_{s-1}+\sum_{j=1}^{k-1}y^j_{(s-1)h+(k-j)n}T^{k-j}_{s-1}+y^k_{(s-1)h}. \label{res4}
\end{eqnarray}
We can solve (\ref{res3}) and (\ref{res4}) to obtain the following.
\begin{pro} \label{pro:1}
A  solution of (\ref{res3}) and (\ref{res4}) with initial condition $T^k_1=y^k_0$ for $k\geq 1$  is given by an infinite number of quasi-homogeneous polynomials 
\begin{equation}
T^k_s=\sum_{K}\sum_{0\leq \lambda_1<\cdots<\lambda_p\leq s-1}y_{\lambda_1h}^{k_1}y_{\lambda_2h+k_1n}^{k_2}\cdots
y_{\lambda_ph+(k_1+\cdots+k_{p-1})n}^{k_p}.
\label{main}
\end{equation}
\end{pro}
Let us remark that the summation in (\ref{main}) is performed over all compositions $K=(k_1,\ldots, k_p)$ of number $k$ with $k_j\geq 1$.

\noindent
\textbf{Proof of proposition \ref{pro:1}.} 
 Firstly, we remark that (\ref{main}) gives $T^k_1=y^k_0$ for all $k\geq 1$. Next, we observe that the polynomials $T^k_s=T^k_s\{y_j^r\}$ are uniquely defined from these  recurrence relations starting from $T^k_1=y^k_0$. Thus, to prove this proposition we have to show that the polynomials $T^k_s$ defined by explicit expression (\ref{main}), solve (\ref{res3}) and (\ref{res4}). To this aim, we consider the partition of the set $D_{p, s}\equiv\{\lambda_j : 0\leq \lambda_1<\cdots<\lambda_p\leq s-1\}$ into two non-intersecting subsets $D_{p, s}^{(1)}\equiv\{\lambda_j : \lambda_1=0; 1\leq \lambda_2<\cdots<\lambda_p\leq s-1\}$ and $D_{p, s}^{(2)}\equiv\{\lambda_j :  1\leq \lambda_1<\cdots<\lambda_p\leq s-1\}$. 
It is evident that
\[
T^{k,h}_{s-1}=\sum_{K}\sum_{\{\lambda_j\}\in D_{p, s}^{(2)}}y_{\lambda_1h}^{k_1}y_{\lambda_2h+k_1n}^{k_2}\cdots
y_{\lambda_ph+(k_1+\cdots+k_{p-1})n}^{k_p}.
\]
Moreover
\begin{eqnarray}
&&\sum_{K}\sum_{\{\lambda_j\}\in D_{p, s}^{(1)}}y_{\lambda_1h}^{k_1}y_{\lambda_2h+k_1n}^{k_2}\cdots
y_{\lambda_ph+(k_1+\cdots+k_{p-1})n}^{k_p} \nonumber \\
&=& \sum_{K}y_{0}^{k_1} \sum_{1\leq \lambda_2<\cdots<\lambda_p\leq s-1}y_{\lambda_2h+k_1n}^{k_2}\cdots
y_{\lambda_ph+(k_1+\cdots+k_{p-1})n}^{k_p}. \nonumber
\end{eqnarray}
It is obvious that the set $\mathcal{K}$ of  all compositions of number $k$  can be presented as $\mathcal{K}=\bigsqcup_{j=1}^k\mathcal{K}_j$, where $\mathcal{K}_j$ stands for the set of compositions of $k$ with $k_1=j$. Clearly, if $K\in\mathcal{K}_j$ then the set $(k_2,\ldots, k_p)$ presents a composition of number $k-j$. Taking this into account, we obtain
\begin{eqnarray}
&&\sum_{K}\left(y_{0}^{k_1}\sum_{1\leq \lambda_2<\cdots<\lambda_p\leq s-1}y_{\lambda_2h+k_1n}^{k_2}\cdots
y_{\lambda_ph+(k_1+\cdots+k_{p-1})n}^{k_p}\right) \nonumber \\
&=&\sum_{j=1}^k\left(y^j_0\sum_{(k_2,\ldots, k_p)}\sum_{1\leq \lambda_2<\cdots<\lambda_p\leq s-1}y_{\lambda_2h+k_1n}^{k_2}\cdots
y_{\lambda_ph+(k_1+\cdots+k_{p-1})n}^{k_p}\right) \nonumber \\
&=&\sum_{j=1}^ky^j_0T^{k-j, h+jn}_{s-1}. \nonumber
\end{eqnarray}
Therefore we proved that (\ref{res3}) is an identity for polynomials defined by (\ref{main}). To prove that (\ref{res4}) is also an identity for (\ref{main}), we use similar reasonings. Therefore, the proposition is proved. \opensquare

Consider now a pseudo-difference operator
\[
\mathcal{S}_1\equiv(\mathcal{T}_1)^{-1}=\Lambda^{h}+\sum_{j\geq 1}(-1)^jz^{-j}S^j(i-(j-1)n)\Lambda^{h-jn}
\]
and its powers
\[
\mathcal{S}_s\equiv(\mathcal{S}_1)^{s}=\Lambda^{sh}+\sum_{j\geq 1}(-1)^jz^{-j}S^j_s(i-(j-1)n)\Lambda^{sh-jn}.
\]
Clearly, each coefficient $S^k_s$ is some discrete polynomial of the fields $\{T^1,\ldots, T^k\}$. Using obvious relations $\mathcal{S}_{s_1-s_2}=\mathcal{S}_{s_1}\mathcal{T}_{s_2}=\mathcal{T}_{s_2}\mathcal{S}_{s_1}$ for $s_1\geq s_2$ we get
\begin{eqnarray}
S^k_{s_1-s_2}(i)&=&S^k_{s_1}(i)+\sum_{j=1}^{k-1}(-1)^jT^j_{s_2}(i+(s_1-s_2)h)S^{k-j}_{s_1}(i+jn) \nonumber \\
&&+(-1)^kT^k_{s_2}(i+(s_1-s_2)h) 	\label{s11} \\
&=&S^k_{s_1}(i-s_2h)+\sum_{j=1}^{k-1}(-1)^jT^j_{s_2}(i-s_2h+(k-j)n)S^{k-j}_{s_1}(i-s_2h)  \nonumber  \\
&&+(-1)^kT^k_{s_2}(i-s_2h) \label{s12}
\end{eqnarray}
and in particular
\begin{eqnarray}
\fl
S^k_{s-1}(i)&=&S^k_s(i)+\sum_{j=1}^{k-1}(-1)^jT^j(i+(s-1)h)S^{k-j}_s(i+jn)+(-1)^kT^k(i+(s-1)h) \label{5} \\
\fl
            &=&S^k_s(i-h)+\sum_{j=1}^{k-1}(-1)^jT^j(i-h+(k-j)n)S^{k-j}_s(i-h)+(-1)^kT^k(i-h). \label{6}
\end{eqnarray}
Let us rewrite (\ref{5}) and (\ref{6}) in equivalent form as relations for multi-variate polynomials $S^k_s\{y^r_j\}$, that is,
\begin{equation}
S^k_{s-1}=S^{k}_{s}+\sum_{j=1}^{k-1}(-1)^jy^j_{(s-1)h}S^{k-j, jn}_{s}+(-1)^ky^k_{(s-1)h}
\label{7}
\end{equation}
and
\begin{equation}
S^{k, h}_{s-1}=S^{k}_{s}+\sum_{j=1}^{k-1}(-1)^jy^j_{(k-j)n}S^{k-j}_{s}+(-1)^ky^k_{0}.
\label{8}
\end{equation}
\begin{pro} \label{th:2}
A quasi-homogeneous solution of (\ref{7}) and (\ref{8}) is given by
\begin{equation}
S^k_s=(-1)^k\sum_{K}(-1)^p\sum_{0\leq \lambda_1\leq\cdots\leq\lambda_p\leq s-1}y_{\lambda_1h+(k_2+\cdots+k_{p})n}^{k_1}\cdots y_{\lambda_{p-1}h+k_pn}^{k_{p-1}}y_{\lambda_ph}^{k_p}.
\label{main1}
\end{equation}
\end{pro}
\textbf{Proof.}  
To prove that (\ref{main1}) satisfy (\ref{8}) we consider a partition of the set $B_{p, s}\equiv\{\lambda_j : 0\leq \lambda_1\leq\cdots\leq\lambda_p\leq s-1\}$ into two non-intersecting subsets $B_{p, s}^{(1)}\equiv\{\lambda_j : \lambda_1=0; 0\leq \lambda_2\leq\cdots\leq\lambda_p\leq s-1\}$ and $B_{p, s}^{(2)}\equiv\{\lambda_j :  1\leq \lambda_1\leq\cdots\leq\lambda_p\leq s-1\}$ what gives
\begin{eqnarray}
\fl
S^k_s&=&(-1)^k\sum_{K}\sum_{\{\lambda_j\}\in B_{p, s}^{(2)}}(-1)^py_{\lambda_1h+(k_2+\cdots+k_{p})n}^{k_1}\cdots y_{\lambda_{p-1}h+k_pn}^{k_{p-1}}y_{\lambda_ph}^{k_p} \nonumber \\
\fl
&&+(-1)^k\sum_{K}(-1)^py_{(k_2+\cdots+k_{p})n}^{k_1}\sum_{0\leq \lambda_2\leq\cdots\leq\lambda_p\leq s-1}y_{\lambda_2h+(k_3+\cdots+k_{p})n}^{k_2}\cdots y_{\lambda_{p-1}h+k_pn}^{k_{p-1}}y_{\lambda_ph}^{k_p}. \nonumber
\end{eqnarray}
The first member in this sum gives $S^{k, h}_{s-1}$, while using the partition $\mathcal{K}=\bigsqcup_{j=1}^k\mathcal{K}_j$ for the second member yields
\[
\sum_{j=1}^{k-1}(-1)^{j+1}y^j_{(k-j)n}S^{k-j}_{s}+(-1)^{k+1}y^k_{0}.
\]
To prove that (\ref{7}) is also an identity for (\ref{main1}), we use similar arguments. Therefore the proposition is proved. \opensquare

Let us remark that in particular case $s=1$, (\ref{main1}) becomes
\begin{equation}
S^k_1=(-1)^k\sum_{K}(-1)^py_{(k_2+\cdots+k_{p})n}^{k_1}\cdots y_{k_pn}^{k_{p-1}}y_{0}^{k_p}
\label{Sk1}
\end{equation}
Let us denote $x^r_j=S^{r,j}_1$. From (\ref{7}) and (\ref{8}) we obtain that
\begin{equation}
x^{k}_{j}+\sum_{q=1}^{k-1}(-1)^qy^{q}_{j+(k-q)n}x^{k-q}_{j}+(-1)^ky^k_{j}=0
\label{9}
\end{equation}
and
\begin{equation}
x^{k}_{j}+\sum_{q=1}^{k-1}(-1)^qy^j_{j}x^{k-j}_{j+qn}+(-1)^ky^k_{j}=0.
\label{10}
\end{equation}
It is evident that (\ref{10}) can be obtained from (\ref{9}) and vice versa by replacing $y^r_j\leftrightarrow x^r_j$. This means that (\ref{Sk1}) gives an invertible involutive transformation which allows one to express $y^r_j$ as a polynomial in $x^r_j$ as
\begin{eqnarray}
y^k_j&=&(-1)^k\sum_{K}(-1)^px_{j+(k_2+\cdots+k_{p})n}^{k_1}\cdots x_{j+k_pn}^{k_{p-1}}x_{j}^{k_p} \nonumber \\
&=&(-1)^k\sum_{K}(-1)^px^{k_1}_jx^{k_2}_{j+k_1n}\cdots x^{k_p}_{j+(k_1+\cdots +k_{p-1})n}. \nonumber 
\end{eqnarray}
Making use of the latter we can express any multi-variate polynomials $Q\{y^r_j\}$ in variables $x^r_j$.
In particular, the following is valid.
\begin{pro}
Polynomials $T^k_s$ and $S^k_s$ are expressed via the set of variables $\{x^r_j\}$ as
\begin{equation}
S^k_s=\sum_{K}\sum_{\{\lambda_j\}\in D_{p, s}}x_{\lambda_1h+(k_2+\cdots+k_{p})n}^{k_1}\cdots x_{\lambda_{p-1}h+k_pn}^{k_{p-1}}x_{\lambda_ph}^{k_p}
\label{Sks}
\end{equation}
and
\begin{equation}
T^k_s=(-1)^k\sum_{K}\sum_{\{\lambda_j\}\in B_{p, s}}(-1)^px_{\lambda_1h}^{k_1}x_{\lambda_2h+k_1n}^{k_2}\cdots
x_{\lambda_ph+(k_1+\cdots+k_{p-1})n}^{k_p}.
\label{Tks}
\end{equation}
\end{pro}

We do not give the proof of this proposition because it is very similar to that of propositions \ref{pro:1} and \ref{th:2}. It is enough to say that
(\ref{Sks}) and (\ref{Tks}) solve the recurrence relations
\begin{eqnarray}
S^k_{s}&=&S^k_{s-1}+\sum_{j=1}^{k-1}x^j_{(s-1)h}S^{k-j, jn}_{s-1}+x^k_{(s-1)h} \label{55} \\
&=&S^{k, h}_{s-1}+\sum_{j=1}^{k-1}x^j_{(k-j)n}S^{k-j, h}_{s-1}+x^k_0 		\label{66}
\end{eqnarray}
and
\begin{equation}
T^k_{s-1}=T^k_s+\sum_{j=1}^{k-1}(-1)^jx^j_{(s-1)h+(k-j)n}T^{k-j}_s+(-1)^kx^k_{(s-1)h} 
\label{555}
\end{equation}
and
\begin{equation}
T^{k, h}_{s-1}=T^{k}_s+\sum_{j=1}^{k-1}(-1)^jx^j_{0}T^{k-j, jn}_s+(-1)^kx^k_{0}, 
\label{666}
\end{equation}
respectively. Thus, one sees that the variables $y^r_j$ and $x^r_j$ are on equal footing.

Putting $s_1=s_2\equiv s$  with arbitrary $s\geq 1$ in (\ref{s11}) and (\ref{s12}) gives the relations
which we rewrite as
\begin{equation}
S^k_{s}+\sum_{j=1}^{k-1}(-1)^jT^j_{s}S^{k-j, jn}_{s}+(-1)^kT^k_{s}=0
\label{skss1}
\end{equation}
and
\begin{equation}
S^k_{s}+\sum_{j=1}^{k-1}(-1)^jT^{j, (k-j)n}_{s}S^{k-j}_{s}+(-1)^kT^k_{s}=0.
\label{skss2}
\end{equation}
Solving these equations yields the following.
\begin{pro}
Polynomials $T^k_s$ and $S^k_s$ are expressed via each other as
\begin{equation}
S^k_s=(-1)^k\sum_{K}(-1)^pT^{k_1}_sT^{k_2, k_1n}_s\cdots T^{k_p, (k_1+\cdots +k_{p-1})n}_s
\label{skss3}
\end{equation}
and
\begin{equation}
T^k_s=(-1)^k\sum_{K}(-1)^pS^{k_1}_sS^{k_2, k_1n}_s\cdots S^{k_p, (k_1+\cdots +k_{p-1})n}_s.
\label{skss4}
\end{equation}
\end{pro}
\noindent
\textbf{Proof.} The proof of this proposition, in fact, follows the lines proposed in \cite{Macdonald} to express the elementary and complete symmetric polynomials via each other in the determinant form with minor modification. We only remark that one can rewrite  (\ref{skss1}) and (\ref{skss2}) as two matrix equations $\mathrm{S}_s\mathrm{T}_s=I$ and $\mathrm{T}_s\mathrm{S}_s=I$, respectively, with semi-infinite matrices
\[
\mathrm{T}_s=\left(
\begin{array}{cccccc}
1 & 0 & 0 & 0 &  0 & \ldots \\
T^1_s & 1 & 0 & 0 &  0 & \ldots \\
T^2_s & T^{1, n}_s & 1 & 0 &  0 & \ldots \\
T^3_s & T^{2, n}_s & T^{1, 2n}_s & 1 &  0 & \ldots \\
\vdots & \vdots & \vdots & \vdots &  \vdots & \vdots
\end{array}
\right)
\]
and
\[
\mathrm{S}_s=\left(
\begin{array}{cccccc}
1 & 0 & 0 & 0 &  0 & \ldots \\
-S^1_s & 1 & 0 & 0 &  0 & \ldots \\
S^2_s & -S^{1, n}_s & 1 & 0 &  0 & \ldots \\
-S^3_s & S^{2, n}_s & -S^{1, 2n}_s & 1 &  0 & \ldots \\
\vdots & \vdots & \vdots & \vdots &  \vdots & \vdots
\end{array}
\right)
\]
and $I$ being semi-infinite identity matrix and use them the known  expression for the elements of an inverse matrix via the elements of the original one. As a result, in particular,  we obtain formulas (\ref{skss3}) and
(\ref{skss4}), which can be presented in the determinant form. \opensquare


\subsection{Polynomials $\tilde{T}^j_{s}$ and $\tilde{S}^j_{s}$}
Given an infinite number of constants $c_j$, define
\[
\tilde{\mathcal{T}}_s\equiv \mathcal{T}_s+\sum_{j\geq 1}z^{-jh}c_j\mathcal{T}_{s+jn}.
\]
Observe that
\[
\tilde{\mathcal{T}}_s=\Lambda^{-sh}+\sum_{j\geq 1}z^{-j}\tilde{T}^j_{s}(i-(j-1)n-sh)\Lambda^{-sh-jn}
\]
with  coefficients $\tilde{T}^j_{s}$ which can be described as follows.
Let $k=\kappa h+r$, where $r$ stands for the remainder of division of $k$ by $h$, then
\[
\tilde{T}^k_s(i)=T^k_s(i)+\sum_{j=1}^{\kappa}c_jT^{k-jh}_{s+jn}(i).
\]
Given an infinite number of constants $H_j$, define
\[
\tilde{\mathcal{S}}_s\equiv \mathcal{S}_s+\sum_{j\geq 1}z^{-jh}H_j\mathcal{S}_{s-jn}.
\]
Here it is supposed that $s-jn\leq 1$. Observe that
\[
\tilde{\mathcal{S}}_s=\Lambda^{sh}+\sum_{j\geq 1}(-1)^jz^{-j}\tilde{S}^j_{s}(i-(j-1)n)\Lambda^{sh-jn}
\]
with
\[
\tilde{S}^k_s(i)=S^k_s(i)+\sum_{j=1}^{\kappa}(-1)^{jh}H_jS^{k-jh}_{s-jn}(i+jhn).
\]
It is evident that
\[
\tilde{\mathcal{T}}_{s_1+s_2}=\tilde{\mathcal{T}}_{s_1}\mathcal{T}_{s_2}=\mathcal{T}_{s_2}\tilde{\mathcal{T}}_{s_1},\;\;
\tilde{\mathcal{S}}_{s_1+s_2}=\tilde{\mathcal{S}}_{s_1}\mathcal{S}_{s_2}=\mathcal{S}_{s_2}\tilde{\mathcal{S}}_{s_1},
\]
\[
\tilde{\mathcal{T}}_{s_1-s_2}=\tilde{\mathcal{T}}_{s_1}\mathcal{S}_{s_2}=\mathcal{S}_{s_2}\tilde{\mathcal{T}}_{s_1},\;\;
\tilde{\mathcal{S}}_{s_1-s_2}=\tilde{\mathcal{S}}_{s_1}\mathcal{T}_{s_2}=\mathcal{T}_{s_2}\tilde{\mathcal{S}}_{s_1}.
\]
This means, in particular, that polynomials $\tilde{T}^k_{s}$ and $\tilde{S}^k_{s}$ as well as   polynomials $T^k_{s}$ and $S^k_{s}$ also satisfy identities (\ref{res1}), (\ref{res2}), (\ref{5}),
(\ref{6}), (\ref{55}), (\ref{66}), (\ref{555}) and (\ref{666}). The difference between these solutions is that in the case of quasi-homogeneous polynomials  $T^k_{s}$ and $S^k_{s}$ one starts from $T^k_1=T^k$ and $S^k_1$ given by (\ref{Sk1}), while for deriving polynomials $\tilde{T}^k_{s}$ and $\tilde{S}^k_{s}$ one uses as the initial data in the corresponding recurrence relations some linear combinations of quasi-homogeneous polynomials.


\subsection{$l$-reduced polynomials}
Requiring $y^{r}_j\equiv 0$, for $r\geq l+1$, we obtain $l$-reduced polynomials $T^k_s$ and $S^k_s$. 
For example, $1$-reduced polynomials look  as
\begin{equation}
T^k_s=\sum_{\{\lambda_j\}\in D_{k, s}}y_{\lambda_1h}y_{\lambda_2h+n}\cdots
y_{\lambda_kh+(k-1)n}
\label{Tks11}
\end{equation}
and
\begin{equation}
S^k_s=\sum_{\{\lambda_j\}\in B_{k, s}}y_{\lambda_1h+(k-1)n}\cdots y_{\lambda_{k-1}h+n}y_{\lambda_kh}
\label{Sks11}
\end{equation}
where $y_j\equiv y^1_j$. These polynomials were reported in \cite{Svinin1}. One can also consider  $l$-reduction of polynomials (\ref{Sks}) and (\ref{Tks}) defined by the condition $x^{r}_j\equiv 0$, for $r\geq l+1$.

One sees that in the degenerate case $n=0$ and $h=1$ (\ref{Tks11}) and (\ref{Sks11}) are nothing but elementary and complete symmetric polynomials, respectively \cite{Macdonald}. We could expect that  polynomials given by (\ref{main}) and (\ref{main1}) keep some nice properties of symmetric ones. In this connection, the following question could be addressed: whether polynomials $T^k_s$  and $S^k_s$ in the general case generate some algebraic structure?  


\subsection{Bibliographical remarks}
It is worth mentioning that $1$-reduced discrete polynomials $T^k_s$ given by (\ref{Tks11}) in the particular case $h=1$ and $n=1$ first appeared in \cite{Tran1}, where they were used for description of the integrals of the mKdV and sine-Gordon discrete equations. In this connection see also \cite{Demskoi}, \cite{Tran1} and \cite{Tran2}. In \cite{Svinin1}, we have shown explicit form of one-field lattice integrable hierarchies including the Itoh-Narita-Bogoyavlenskii lattice hierarchy. In \cite{Svinin6} we  derived a class of lattice integrable hierarchies related by some Miura-type transformation to the Itoh-Narita-Bogoyavlenskii lattice. We remark that this class of hierarchies  contains in particular Yamilov lattice hierarchy
\cite{Yamilov}.

We believe that multi-variate polynomials $T^k_s$ and $S^k_s$ given by (\ref{main}) and (\ref{main1}) and their $l$-reduced versions  could be widely applicable in the theory of classical integrable systems. In the following two sections we are going to support this point by constructing  discrete equations having a Lax representation in terms of these polynomials and showing then the sense of these discrete equations from the point of view of lattice integrable hierarchies.


\section{Discrete equations admitting a Lax pair representation}
\label{sect2}
This section is designed to show infinite and finite systems of discrete equations which share the property of having Lax pair representation.
We define our discrete systems with the help of some generating equations on formal pseudo-difference operators.


\subsection{The first class of discrete equations}
Consider a partition $\mathcal{T}_s=(\mathcal{T}_s)_{+,k}+(\mathcal{T}_s)_{-,k}$, where
\[
(\mathcal{T}_s)_{+,k}\equiv\Lambda^{-sh}+\sum_{j=1}^kz^{-j}T^j_{s}(i-(j-1)n-sh)\Lambda^{-sh-jn}
\]
and
\[
(\mathcal{T}_s)_{-,k}\equiv\sum_{j> k}z^{-j}T^j_{s}(i-(j-1)n-sh)\Lambda^{-sh-jn}.
\]
Observe that
\[
[(\mathcal{T}_s)_{+,k}, \mathcal{T}_1]=\sum_{j\geq 1}z^{-k-j}\nu_j(i)\Lambda^{-(s+1)h-(k+j)n},
\]
where
\begin{eqnarray}
\fl
\nu_j(i)&=&T^{k+j}(i-(s+1)h-(k+j-1)n) \nonumber \\
\fl
&&+\sum_{q=1}^kT^q_s(i-sh-(q-1)n)T^{k+j-q}(i-(s+1)h-(k+j-1)n) \nonumber \\
\fl
&&-T^{k+j}(i-h-(k+j-1)n) \nonumber \\
\fl
&&-\sum_{q=1}^kT^q_s(i-(s+1)h-(k+j-1)n)T^{k+j-q}(i-h-(k+j-q-1)n). \nonumber
\end{eqnarray}
Let
\[
(\tilde{\mathcal{T}}_s)_{+,k}=(\mathcal{T}_s)_{+,k}+\sum_{j\geq 1}z^{-jh}c_j(\mathcal{T}_{s+jn})_{+,k-jh}.
\]
As can be checked
\[
(\tilde{\mathcal{T}}_s)_{+,k}=\Lambda^{-sh}+\sum_{j=1}^{k}z^{-j}\tilde{T}^j_{s}(i-(j-1)n-sh)\Lambda^{-sh-jn}
\]
and the commutativity equation
\begin{equation}
[(\tilde{\mathcal{T}}_s)_{+,k}, \mathcal{T}_1]=0
\label{comm1}
\end{equation}
generate an infinite number of discrete equations
\[
T^{k+j}(i+sh)+\sum_{q=1}^k\tilde{T}^q_{s}(i)T^{k+j-q}(i+sh+qn)
\]
\begin{equation}
=T^{k+j}(i)+\sum_{q=1}^k\tilde{T}^q_{s}(i+h+(k+j-q)n)T^{k+j-q}(i),\;\;\; j\geq 1.
\label{disc1}
\end{equation}
Clearly, this system admits an $l$-reduction with the help of condition $T^j\equiv 0$ for $j\geq l+1$ yielding therefore an $l$-field system of ordinary difference equations.

Commutativity equation (\ref{comm1}) appears as a condition of the compatibility of two linear equations
\[
(\tilde{\mathcal{T}}_s)_{+,k}(\psi_i)=\frac{w}{z^k}\psi_i,\;\;\;
\mathcal{T}_1(\psi_i)=\psi_i
\]
the first of which we can rewrite as
\begin{equation}
w\psi_{i+kn+sh}=z^k\psi_{i+kn}+\sum_{j=1}^kz^{k-j}\tilde{T}^j_{s}(i+(k-j+1)n)\psi_{i+(k-j)n}.
\label{lin333}
\end{equation}
Remark that  $w$ here is the second spectral parameter. The equation  $\mathcal{T}_1(\psi_i)=\psi_i$ with $l$-reduced operator $\mathcal{T}_1$ looks as
\begin{equation}
z^l\psi_{i+ln+h}=z^l\psi_{i+ln}+\sum_{j=1}^lz^{l-j}T^j(i+(l-j+1)n)\psi_{i+(l-j)n}.
\label{l1}
\end{equation}

As a consequence of (\ref{comm1}) we have
\[
[(\tilde{\mathcal{T}}_s)_{+,k}, \mathcal{S}_1]=0
\]
which gives an infinite number of discrete equations
\[
S^{k+j}(i+sh)+\sum_{q=1}^k(-1)^q\tilde{T}^q_{s}(i+h)S^{k+j-q}(i+sh+qn)
\]
\begin{equation}
=S^{k+j}(i)+\sum_{q=1}^k(-1)^q\tilde{T}^q_{s}(i+(k+j-q)n)S^{k+j-q}(i),\;\;\; j\geq 1.
\label{disc2}
\end{equation}
This system also admits an $l$-reduction with the help of condition $S^j\equiv 0$ for $j\geq l+1$. Linear problem $\mathcal{S}_1(\psi_i)=\psi_i$, in this case, becomes
\begin{equation}
z^l\psi_{i+ln-h}=z^l\psi_{i+ln}+\sum_{j=1}^l(-1)^jz^{l-j}S^j(i+(l-j+1)n-h)\psi_{i+(l-j)n}.
\label{l2}
\end{equation}


\subsection{The second class of discrete equations}
Consider a partition $\mathcal{S}_s=(\mathcal{S}_s)_{+,k}+(\mathcal{S}_s)_{-,k}$, where
\[
(\mathcal{S}_s)_{+,k}\equiv\Lambda^{sh}+\sum_{j=1}^k(-1)^jz^{-j}S^j_{s}(i-(j-1)n)\Lambda^{sh-jn}
\]
and
\[
(\mathcal{S}_s)_{-,k}\equiv\sum_{j> k}(-1)^jz^{-j}S^j_{s}(i-(j-1)n)\Lambda^{ssh-jn}.
\]
Observe that
\begin{equation}
[(\mathcal{S}_s)_{+,k}, \mathcal{T}_1]=\sum_{j\geq 1}z^{-k-j}\mu_j(i)\Lambda^{(s-1)h-(k+j)n},
\label{observe}
\end{equation}
with
\begin{eqnarray}
\fl
\mu_j(i)&=&T^{k+j}(i+(s-1)h-(k+j-1)n) \nonumber \\
\fl
&&+\sum_{q=1}^k(-1)^qS^q_s(i-(q-1)n)T^{k+j-q}(i+(s-1)h-(k+j-1)n) \nonumber \\
\fl
&&-T^{k+j}(i-h-(k+j-1)n) \nonumber \\
\fl
&&-\sum_{q=1}^k(-1)^qS^q_s(i-h-(k+j-1)n)T^{k+j-q}(i-h-(k+j-q-1)n). \nonumber
\end{eqnarray}
Let
\[
(\tilde{\mathcal{S}}_s)_{+,k}\equiv  (\mathcal{S}_s)_{+,k}+\sum_{j\geq 1}z^{-jh}H_j(\mathcal{S}_{s-jn})_{+, k-jh}
\]
then
\[
(\tilde{\mathcal{S}}_s)_{+,k}=\Lambda^{sh}+\sum_{j=1}^k(-1)^jz^{-j}\tilde{S}^j_s(i-(j-1)n)\Lambda^{sh-jn}.
\]
We define the second class of discrete equations with the help of commutativity equation
\begin{equation}
[(\tilde{\mathcal{S}}_s)_{+,k}, \mathcal{T}_1]=0,
\label{comm2}
\end{equation}
which yields an infinite number of discrete equations
\[
T^{k+j}(i+sh)+\sum_{q=1}^k(-1)^q\tilde{S}^q_{s}(i+h+(k+j-q)n)T^{k+j-q}(i+sh)
\]
\begin{equation}
=T^{k+j}(i)+\sum_{q=1}^k(-1)^q\tilde{S}^q_{s}(i)T^{k+j-q}(i+qn),\;\;\; j\geq 1.
\label{disc3}
\end{equation}

As a consequence of (\ref{comm2}) we have Lax equation
\begin{equation}
[(\tilde{\mathcal{S}}_s)_{+,k}, \mathcal{S}_1]=0,
\label{comm3}
\end{equation}
which gives
\[
S^{k+j}(i+sh)+\sum_{q=1}^k\tilde{S}^q_{s}(i+(k+j-q)n)S^{k+j-q}(i+sh)
\]
\begin{equation}
=S^{k+j}(i)+\sum_{q=1}^k\tilde{S}^q_{s}(i+h)S^{k+j-q}(i+qn),\;\;\; j\geq 1.
\label{disc4}
\end{equation}

Clearly, commutativity equations (\ref{comm2}) and (\ref{comm3}) appear as a conditions of the consistency of the linear equation
\[
(\tilde{\mathcal{S}}_s)_{+,k}(\psi_i)=\frac{w}{z^k}\psi_i
\]
which we can rewrite as
\begin{equation}
w\psi_{i-sh+kn}=z^k\psi_{i+kn}+\sum_{j=1}^k(-1)^jz^{k-j}\tilde{S}^j_{s}(i-sh+(k-j+1)n)\psi_{i+(k-j)n}
\label{lin222}
\end{equation}
with $\mathcal{T}_1(\psi_i)=\psi_i$ and $\mathcal{S}_1(\psi_i)=\psi_i$, respectively. 


\subsection{Remarks on Lax pairs}
We have constructed  infinite families of ordinary difference systems (\ref{disc1}), (\ref{disc2}), (\ref{disc3}) and (\ref{disc4}), which are equivalent to the consistency conditions of the corresponding pairs of linear equations. We suppose that $\tilde{S}^k_s$ and $\tilde{T}^k_s$ are expressed via the set of the fields $\{T^j\}$  in equations (\ref{disc1}) and (\ref{disc3}), while in (\ref{disc2}) and (\ref{disc4}) they are expressed via $\{S^j\}$. Let us remember that these two sets of dynamical variables are related to each other by involutive invertible transformation. In particular, we are interested in $l$-reduced versions of these systems which are defined by condition $T^j\equiv 0,\; j\geq l+1$ for (\ref{disc1}) and (\ref{disc3})
and $S^j\equiv 0,\; j\geq l+1$ for (\ref{disc2}) and (\ref{disc4}). The first question which naturally arise is: how do we construct the first integrals of the system under consideration  using corresponding pair of linear equations? 

Consider a pair of the linear equations
\[
\psi_{i+N_1}=L_1(\psi_i,\ldots, \psi_{i+N_1-1})
\]
and
\[
\psi_{i+N_2}=L_2(\psi_i,\ldots, \psi_{i+N_2-1}),
\]
for which the compatibility condition is equivalent to some system of ordinary difference equations.  For example, in the case (\ref{lin333}) and (\ref{l1}), we have $N_1=kn+sh$ and $N_2=ln+h$. Choosing any of these equations, say the first one, we construct the vector-function $\Psi(i)=(\psi^{(0)}(i),\ldots, \psi^{(N_1-1)}(i))^T$, where $\psi^{(j)}(i)\equiv \psi_{i+j}$. Then we use the rest equation to construct linear equation $L(i)\Psi(i)=0$ with some matrix-function $L$. The second equation of the form $\Psi(i+1)=A(i)\Psi(i)$ is easily defined by 
\[
\psi^{(j)}(i+1)=\psi^{(j+1)}(i),\;\; j=1,\ldots, N_1-2
\]
and
\[
\psi^{(N_1-1)}(i+1)=L_1(\psi^{(0)}(i),\ldots, \psi^{(N_1-1)}(i)).
\]
Therefore the system under consideration are equivalent to the Lax equation 
\[
L(i+1)A(i)=A(i)L(i). 
\]
The condition $\det L=0$ gives us the spectral curve $P(w, z)=0$.


\subsection{Integrable hierarchies of lattice evolution equations}
From (\ref{observe}) we see that
\[
[(\mathcal{S}_{sn})_{+,sh}, \mathcal{T}_1]=\sum_{j\geq 1}z^{-sh-j}\mu_j(i)\Lambda^{-h-jn}.
\]
This means that we can correctly define evolution equations by generating relation
\begin{equation}
\partial_s\mathcal{T}_1=z^{sh}[(\mathcal{S}_{sn})_{+,sh}, \mathcal{T}_1]
\label{Lax1}
\end{equation}
with
\[
(\mathcal{S}_{sn})_{+,sh}=\Lambda^{shn}+\sum_{j=1}^{sh}(-1)^jz^{-j}S^j_{sn}(i-(j-1)n)\Lambda^{shn-jn}.
\]
This generating equation gives an infinite number of evolution differential-difference equations
\begin{eqnarray}
\fl
\partial_sT^j(i)&=&T^{sh+j}(i)+\sum_{q=1}^{sh}(-1)^qS^q_{sn}(i+h+(j-q)n)T^{sh+j-q}(i) \nonumber \\
\fl
&&-T^{sh+j}(i-shn)-\sum_{q=1}^{sh}(-1)^qS^q_{sn}(i-shn)T^{sh+j-q}(i-(sh-q)n). \label{evolution111}
\end{eqnarray}
As a consequence of Lax equation (\ref{Lax1}) we have
\[
\partial_s\mathcal{S}_1=z^{sh}[(\mathcal{S}_{sn})_{+,sh}, \mathcal{S}_1]
\]
which yields
\begin{eqnarray}
\fl
\partial_sS^j(i)&=&(-1)^{sh}\left\{S^{sh+j}(i)+\sum_{q=1}^{sh}S^q_{sn}(i+(j-q)n)S^{sh+j-q}(i)\right\} \nonumber \\
\fl
&&-(-1)^{sh}\left\{S^{sh+j}(i-shn)     \phantom{\sum_{q=1}^{sh}}              \right.  \nonumber \\
\fl
&&\left. +\sum_{q=1}^{sh}S^q_{sn}(i-shn+h)S^{sh+j-q}(i-(sh-q)n)\right\}. \label{evolution2}
\end{eqnarray}

Observe that
\[
\left\{\partial_s+\sum_{q=1}^{s-1}H_q\partial_{s-q}\right\}\mathcal{T}_1=z^{sh}[(\tilde{\mathcal{S}}_{sn})_{+,sh}, \mathcal{T}_1].
\]
From the latter we see that the stationarity equation
\[
\left\{\partial_s+\sum_{q=1}^{s-1}H_q\partial_{s-q}\right\}\mathcal{T}_1=0
\]
is equivalent to a special case of (\ref{comm2}).


\subsection{Some examples of lattice integrable hierarchies}
Using (\ref{evolution111}) and (\ref{evolution2}), we are able to present the explicit form for a number of lattice integrable hierarchies.


\subsubsection{Itoh-Narita-Bogoyavlenskii lattice hierarchy}
Let $h=1$ while $n\geq 1$ is an arbitrary integer. Lax equation (\ref{Lax1}) with the operator $\mathcal{T}_1=\Lambda^{-1}+z^{-1}T(i-1)\Lambda^{-1-n}$, where $T(i)\equiv T^1(i)$,
gives evolution equations
\begin{equation}
\partial_sT(i)=(-1)^sT(i)\left\{S^s_{sn}(i-(s-1)n+1)-S^s_{sn}(i-sn)\right\}
\label{INB}
\end{equation}
constituting an integrable hierarchy for the extended Volterra equation \cite{Narita}
\[
\partial_1T(i)=T(i)\left\{\sum_{j=1}^nT(i-j)-\sum_{j=1}^nT(i+j)\right\}
\]
also known  as the Itoh-Narita-Bogoyavlenskii lattice \cite{Itoh}, \cite{Narita}, \cite{Bogoyavlenskii}.


\subsubsection{Bogoyavlenskii lattice hierarchy}
Now let $n=1$ while $h\geq 1$ is an arbitrary integer. Lax equation (\ref{Lax1}) with the operator $\mathcal{T}_1=\Lambda^{-h}+z^{-1}T_{i-h}\Lambda^{-h-1}$ generates integrable hierarchy
\[
\partial_sT(i)=(-1)^{sh}T(i)\left\{S^{sh}_{s}(i-(s-1)h+1)-S^{sh}_{s}(i-sh)\right\}
\]
for the Bogoyavlenskii lattice
\[
\partial_1T(i)=(-1)^hT(i)\left\{\prod_{j=1}^hT(i+j)-\prod_{j=1}^hT(i-j)\right\}.
\]


\subsubsection{Lattice Gel'fand-Dickii or  $(l+1)$-KdV hierarchy}
Let $h=1$ and $n=1$, while $l\geq 1$ be an arbitrary integer. With the operator
\[
\mathcal{T}_1=\Lambda^{-1}+\sum_{j=1}^lz^{-j}T^j(i-j)\Lambda^{-1-j}
\]
we obtain the evolution equations
\begin{eqnarray}
\partial_sT^j(i)&=&(-1)^{s+j}\left\{\sum_{q=j}^l(-1)^qS^{s+j-q}_s(i+q-s+1)T^q(i)\right\} \nonumber \\
&&-(-1)^{s+j}\left\{\sum_{q=j}^l(-1)^qS^{s+j-q}_s(i-s)T^q(i+j-q)\right\} \nonumber
\end{eqnarray}
for the lattice $(l+1)$-KdV hierarchy \cite{Antonov}, \cite{Belov1}, \cite{Belov2}. In the case $l=1$, we obtain the Volterra lattice hierarchy
\[
\partial_sT(i)=(-1)^sT(i)\left\{S^s_{s}(i-s+2)-S^s_{s}(i-s)\right\}
\]
which, as is known, gives one of the lattice analogues of the KdV hierarchy. In the case $l=2$, we obtain the lattice Boussinesq hierarchy in the form
\begin{eqnarray}
\partial_sT^1(i)&=&(-1)^s\left\{S^{s}_s(i-s+2)T^1(i)-S^{s-1}_s(i-s+3)T^2(i)\right\} \nonumber \\
&&-(-1)^s\left\{S^{s}_s(i-s)T^1(i)-S^{s-1}_s(i-s)T^2(i-1)\right\}, \nonumber
\end{eqnarray}
\[
\partial_sT^2(i)=(-1)^sT^2(i)\left\{S^{s}_s(i-s+3)-S^{s}_s(i-s)\right\}.
\]
whose the first flow is given by \cite{Belov1}
\[
\partial_1T^1(i)=T^2(i)-T^2(i-1)+T^1(i)\left\{T^1(i-1)-T^1(i+1)\right\},
\]
\[
\partial_1T^2(i)=T^2(i)\left\{T^1(i-1)-T^1(i+2)\right\}.
\]

\subsection{Bibliographical remarks}

We have not  discussed  above a quite important question concerning the Hamiltonian, and what is more important from different points of view, the bi-Hamiltonian structure of lattice integrable hierarchies with a finite and infinite number of fields given by (\ref{evolution111}) and (\ref{evolution2}). In the previous subsection we have only presented some examples of lattice integrable hierarchies known in the literature in their explicit form. One can find studies on the Hamiltonian approach to lattice systems, for example, in \cite{Antonov}, \cite{Belov1} - \cite{Blaszak}, \cite{Bonora}, \cite{Carlet} and \cite{Xiong}.


\section{Integrable hierarchies of lattice evolution equations from the point of view of the KP hierarchy}
\label{sect3}

In the previous  section we have shown some classes of discrete equations with their Lax representation in terms of  polynomials $T^k_s$ and $S^k_s$ which were given in explicit form by (\ref{main}), (\ref{main1}), (\ref{Sks}) and (\ref{Tks}). In this section we have a look at that from another point of view.
In \cite{Svinin2}-\cite{Svinin1} we developed an approach which relates some evolutionary lattice equations and their hierarchies to the KP hierarchy. We state in this section this approach. This aims, in particular, to clarify the appearance of $l$-reduced versions of discrete equations (\ref{disc1}), (\ref{disc2}), (\ref{disc3}) and (\ref{disc4}).


\subsection{Free sequences of KP hierarchies. Invariant submanifolds}
To begin with, we consider the free sequences of KP hierarchies $\{h(i, z) : i\in\mathbb{Z}\}$ in the form of local conservation laws \cite{Cherednik}, \cite{Wilson}, that is,
\[
\partial_sh(i, z)=\partial H^{(s)}(i, z),\;\;\; \forall i\in\mathbb{Z},
\]
where the formal Laurent series $h(z)=z+\sum_{j\geq 2}h_jz^{-j+1}$ and $H^{(s)}(z)=z^s+\sum_{j\geq 1}H_j^sz^{-j}$ are related to the KP wave function by $h(z)=\partial \psi/\psi$ and $H^{(s)}(z)=\partial_s \psi/\psi$, respectively. Therefore, each coefficient $H_j^s$ is supposed to be some differential polynomial $H_j^s=H_j^s[h_2,\ldots, h_{s+k}]$. More exactly, the Laurent series $H^{(s)}(z)$ is calculated as a projection of $z^s$ on the linear space ${\mathcal H}_{+}=\left\langle 1, h, h^{(2)},...\right\rangle$ spanned by Fa\`a di Bruno iterates $h^{(k)}\equiv (\partial+h)^{k}(1)$ . For example,
\[
H^{(1)}=h,\;\;\;
H^{(2)}=h^{(2)}-2h_2,\;\;\;H^{(3)}=h^{(3)}-3h_2h-3h_3-3h_2^{\prime}
\]
etc. Remark, that $h^{(s)}(z)=\partial^s \psi/\psi$. As is known, dynamics of the coefficients of  $H_j^s$ in virtue of KP flows is defined by the invariance property
\begin{equation}
\left(\partial_s+H^{(s)}\right){\mathcal H}_{+}\subset{\mathcal H}_{+},\;\;\;
\forall s\geq 1.
\label{eq3.0}
\end{equation}
Conversely,  it can be seen, that a hierarchy of dynamical systems governed by (\ref{eq3.0}), known as the Central System, is equivalent to the KP hierarchy \cite{Casati}. Moreover, we consider also a Laurent series
$a(i, z)=z+\sum_{j\geq 1}a_j(i)z^{-j+1}\equiv z\psi_{i+1}/\psi_i$. By its definition, it satisfies the following evolution equations
\begin{equation}
\partial_sa(i, z)=a(i, z)\left\{H^{(s)}(i+1, z)-H^{(s)}(i, z)\right\},
\label{a}
\end{equation}
which in turn can be presented in the form of differential-difference conservation laws
\[
\partial_s\xi(i, z)=H^{(s)}(i+1, z)-H^{(s)}(i, z)
\]
with
\[
\displaystyle
\xi(i, z)=\ln a(i)=\ln z-\sum_{j\geq 1}\frac{1}{j}\left(-\sum_{k\geq 1}a_k(i)z^{-k}\right)^j
\equiv\ln z+\sum_{j\geq 1}\xi_j(i)z^{-j}.
\]
It is convenient to consider a free bi-infinite sequence of KP hierarchies as two equations \cite{Magri}
\begin{equation}
\partial_sh_k(i)=\partial H^s_{k-1}(i),\;\;\;
\partial_s\xi_k(i)=H^s_k(i+1)-H^s_k(i).
\label{eq3.1}
\end{equation}
The two theorems given below yield an infinite number of invariant submanifolds of the phase space of the sequence of KP hierarchies whose points are defined by $\{a_k(i), H^s_k(i) : i\in{\mathbb Z}\}$.
\begin{thm} \cite{Svinin2}
\label{t3.1}
Submanifold  ${\mathcal S}_{k-1}^n$ defined by the condition
\begin{equation}
z^{k-n}a^{[n]}(i, z)\in{\mathcal H}_{+}(i),\;\; \forall i\in{\mathbb Z}
\label{eq3.2}
\end{equation}
is invariant with respect to the flows given by evolution equations (\ref{eq3.1}).
\end{thm}
Here, by definition,
\[
a^{[s]}(i, z)=\left\{
\begin{array}{l}
\prod_{j=1}^sa(i+j-1, z),\;\;\;\mbox{for}\;\;\; s\geq 1; \\
1,\;\;\;\mbox{for}\;\;\; s=0;  \\
\prod_{j=1}^{|s|}a^{-1}(i-j, z),\;\;\;\mbox{for}\;\;\; s\leq -1.
\end{array}
\right.
\]
The condition (\ref{eq3.2}) can be ultimately rewritten  as a single generating relation\footnote{Note that we sometimes do not indicate dependence on discrete variable $i$ in formulas which contain no shifts with respect to this variable.}
\begin{equation}
z^{k-n}a^{[n]}=H^{(k)}+\sum_{j=1}^ka_j^{[n]}H^{(k-j)},
\label{eq3.3}
\end{equation}
where $a_j^{[s]}$ are defined as the coefficients of Laurent series
$a^{[s]}=z^s+\sum_{j\geq 1}a_j^{[s]}z^{s-j}$ being ultimately some quasi-homogeneous polynomials $a_j^{[s]}=a_j^{[s]}[a_1,\ldots, a_j]$.
\begin{thm}\cite{Svinin3}
\label{t3.2}
The following chain of inclusions
\begin{equation}
{\mathcal S}_{s-1}^n\subset {\mathcal S}_{2s-1}^{2n}\subset{\mathcal
S}_{3s-1}^{3n}\subset\cdot\cdot\cdot
\label{eq3.4}
\end{equation}
is valid.
\end{thm}

\subsection{The $n$th discrete KP hierarchy}
Let us consider the restriction of (\ref{eq3.1}) on ${\mathcal S}_0^n$ which is defined by the condition $z^{1-n}a^{[n]}=h+a_1^{[n]}$. Clearly, by this constraint we should get as a result some differential-difference equations on infinite number of dynamical variables $a_k=a_k(i)$ in the form of differential-difference conservation laws, but there is a need to know how coefficients $H^s_j$ are expressed via $a_j$ for all $s\geq 1$. By our assumption, we already know that $H^1_k\equiv h_{k+1}=a_{k+1}^{[n]}$. To obtain explicit expressions for all $H^s_k$ we must  use theorem \ref{t3.2}. For $s=1$, we have a chain
\[
{\mathcal S}_0^n\subset {\mathcal S}_1^{2n}\subset{\mathcal
S}_2^{3n}\subset\cdot\cdot\cdot
\]
Solving, step-by-step, the defining equations for all invariant submanifolds in this chain we come to \cite{Svinin3}
\begin{equation}
H^s_k=F_k^{(n,s)}[a_1,\ldots, a_{k+s}]\equiv a_{k+s}^{[sn]}+\sum_{j=1}^{s-1}q_j^{(n, sn)}a_{k+s-j}^{[(s-j)n]}.
\label{hs}
\end{equation}
Here $q_j^{(n, r)}$, being some quasi-homogeneous polynomials, are defined through the relation
\begin{equation}
z^r=a^{[r]}+\sum_{j\geq 1}q_j^{(n, r)}z^{j(n-1)}a^{[r-jn]}.
\label{aq}
\end{equation}
We can code all relations (\ref{hs}) in one generating relation
\begin{equation}
H^{(s)}=F^{(n, s)}=z^{s-sn}\left\{a^{[sn]}+\sum_{j=1}^sz^{j(n-1)}q_j^{(n, sn)}a^{[(s-j)n]}\right\},
\label{zln}
\end{equation}
where $F^{(n, s)}\equiv z^s+\sum_{j\geq 1}F_j^{(n, s)}z^{-j}$. Thus the restriction of dynamics given by (\ref{eq3.1}) on $\mathcal{S}_0^n$ yields evolution equations in the form of differential-difference conservation laws
\begin{equation}
\partial_s\xi_k(i)=F_k^{(n,s)}(i+1)-F_k^{(n,s)}(i)
\label{dKP}
\end{equation}
on an infinite number of fields $\{a_k=a_k(i)\}$.  Let $\mathcal{M}$ be the corresponding phase space. We call the hierarchy of the flows on $\mathcal{M}$ given by evolution equations (\ref{dKP}) the $n$th discrete KP hierarchy. Let us observe that in terms of KP wave function the relation (\ref{zln}) takes the form
\begin{equation}
\partial_s\psi_i=z^s\psi_{i+sn}+\sum_{j=1}^sz^{s-j}q_j^{(n, sn)}(i)\psi_{i+(s-j)n},
\label{evolution}
\end{equation}
while (\ref{aq}) is equivalent to
\[
z^r\psi_i=z^r\psi_{i+r}+\sum_{j\geq 1}z^{r-j}q_j^{(n, r)}(i)\psi_{i+r-jn}.
\]
As a condition of consistency of these linear equations we obtain the evolution equations
\begin{eqnarray}
\partial_sq_k^{(n, r)}(i)&=&q_{k+s}^{(n, r)}(i+sn)+\sum_{j=1}^sq_j^{(n,sn)}(i)q_{k+s-j}^{(n, r)}(i+(s-j)n)- \nonumber \\
& &-q_{k+s}^{(n, r)}(i)-\sum_{j=1}^sq_j^{(n,sn)}(i+r-(k+s-j)n)q_{k+s-j}^{(n, r)}(i) \label{dts}
\end{eqnarray}
which, by their construction, are identities by virtue of the $n$th discrete KP hierarchy (\ref{dKP}).


\subsection{Some identities for $q_j^{(n, s)}$ and $a_j^{[s]}$}
Relation (\ref{aq}) yields the infinite triangle system
\begin{equation}
a_k^{[r]}+\sum_{j=1}^{k-1}a^{[r-jn]}_{k-j}q_j^{(n,r)}+q_k^{(n,r)}=0,\;\;\; k\geq 1
\label{triangle}
\end{equation}
from which we can calculate, for example,
\[
q_1^{(n, r)}=-a_1^{[r]},\;\;\; q_2^{(n, r)}=-a_2^{[r]}+a_1^{[r]}a_1^{[r-n]},\;\;\;
\]
\[
q_3^{(n, r)}=-a_3^{[r]}+a_1^{[r]}a_2^{[r-n]}+a_1^{[r-2n]}a_2^{[r]}-a_1^{[r]}a_1^{[r-n]}a_1^{[r-2n]}
\]
etc. It was shown in \cite{Svinin1} that if (\ref{triangle}) is valid then a more general identity
\begin{equation}
a_k^{[p]}(i)=a_k^{[r]}(i)+\sum_{j=1}^{k-1}a_{k-j}^{[r-jn]}(i)q_j^{(n,r-p)}(i+p)+q_k^{(n,r-p)}(i+p),
\label{aqg}
\end{equation}
for all integers $r$ and $p$, holds. Resolving the latter in favor of $q_k^{(n,r-p)}(i+p)$ yields
\[
q_k^{(n,r-m)}(i+m)=a_k^{[m]}(i)+\sum_{j=1}^{k-1}q_j^{(n,r-(k-j)n)}(i)a_{k-j}^{[m]}(i)+q_k^{(n,r)}(i).
\]
We remark that polynomials $a_j^{[s]}$, by definition, must satisfy
\begin{eqnarray}
a_k^{[s_1+s_2]}(i)&=&a_k^{[s_1]}(i)+\sum_{j=1}^{k-1}a_j^{[s_1]}(i)a_{k-j}^{[s_2]}(i+s_1)+a_k^{[s_2]}(i+s_1) \nonumber \\
&=&a_k^{[s_2]}(i)+\sum_{j=1}^{k-1}a_j^{[s_2]}(i)a_{k-j}^{[s_1]}(i+s_2)+a_k^{[s_1]}(i+s_2) \label{r1r2}
\end{eqnarray}
while for $q_j^{(n, s)}$ we have
\begin{eqnarray}
q_k^{(n, s_1+s_2)}(i)&=&q_k^{(n, s_1)}(i)+\sum_{j=1}^{k-1}q_j^{(n, s_1)}(i)q_{k-j}^{(n, s_2)}(i+s_1-jn)+q_k^{(n, s_2)}(i+s_1) \nonumber \\
&=&q_k^{(n, s_2)}(i)+\sum_{j=1}^{k-1}q_j^{(n, s_2)}(i)q_{k-j}^{(n, s_1)}(i+s_2-jn)+q_k^{(n, s_1)}(i+s_2). \nonumber
\end{eqnarray}


\subsection{Restriction of the $n$th discrete KP hierarchy on submanifold ${\mathcal M}_{n, p, l}$}
To obtain in this framework the description of some integrable differential-difference systems with a finite number of fields, we consider the intersection ${\mathcal S}_0^n\cap{\mathcal S}_{l-1}^p$. Obviously, it is equivalent to the restriction of the flows of the $n$th discrete KP hierarchy on a submanifold ${\mathcal M}_{n, p, l}\subset{\mathcal M}$. Taking into account (\ref{eq3.3}) and (\ref{hs}), we immediately obtain that defining equations for ${\mathcal M}_{n, p, l}$ are coded in the relation
\[
z^{l-p}a^{[p]}=F^{(n, l)}+\sum_{j=1}^la_j^{[p]}F^{(n, l-j)}
\]
from which we derive
\begin{equation}
J_k^{(n, p, l)}[a_1,\ldots, a_{k+l}]=0,\;\;\; \forall k\geq 1
\label{mnpl}
\end{equation}
with
\begin{eqnarray}
J_k^{(n, p, l)}&=&a^{[p]}_{k+l}-F_k^{(n, l)}-\sum_{j=1}^{l-1}a_j^{[p]}F_k^{(n, l-j)}  \nonumber \\
               &=&a^{[p]}_{k+l}(i)-a_{k+l}^{[ln]}(i)-\sum_{j=1}^{l-1}q_j^{(n,ln-p)}(i+p)a_{k+l-j}^{[(l-j)n]}(i). \label{jknpl}
\end{eqnarray}
We see from (\ref{jknpl}) that $J_k^{(n, ln, l)}=0$ is just an identity and, therefore, produces no invariant submanifold. This is because ${\mathcal S}_0^n\subset{\mathcal S}_{l-1}^{ln}$ by virtue of  theorem \ref{t3.2}. 

Let us observe that, taking into account (\ref{aqg}), we can rewrite (\ref{jknpl}) as
\begin{equation}
J_k^{(n, p, l)}=Q_k^{(n, p, l)}+\sum_{j=1}^{k-1}a_j^{[-(k-j)n]}Q_{k-j}^{(n, p, l)},
\label{JQ}
\end{equation}
where $Q_k^{(n, p, l)}(i)\equiv q_{k+l}^{(n,ln-p)}(i+p)$. Resolving this quasi-linear infinite triangle system in favor of $Q_k^{(n, p, l)}$ gives
\begin{equation}
Q_k^{(n, p, l)}=J_k^{(n, p, l)}+\sum_{j=1}^{k-1}q_j^{(n, -(k-j)n)}J_{k-j}^{(n, p, l)}
\label{QJ}
\end{equation}
for all $k\geq 1$. This means that, in principle, we can define the submanifold ${\mathcal M}_{n, p, l}$ by equations $Q_k^{(n, p, l)}[a_1,\ldots, a_{k+l}]=0$.


\subsection{Linear equations on KP wave function corresponding to submanifold ${\mathcal M}_{n, p, l}$}
Some remarks about linear equations on the KP formal wave function $\psi_i$ which follows as a result of the restriction of the $n$th discrete KP hierarchy on the submanifold ${\mathcal M}_{n, p, l}$ are in order. Let $J^{(n, p, l)}(z)\equiv\sum_{j\geq 1}J_j^{(n, p, l)}z^{-j}$. We observe that an infinite number of equations (\ref{mnpl}) can be presented by single a generating relation
\[
J^{(n, p, l)}(i, z)=z^{l-p}a^{[p]}(i, z)
\]
\[
-z^{l-ln}\left\{a^{[ln]}(i, z)+\sum_{j=1}^lz^{j(n-1)}q_j^{(n, ln-p)}(i+p)a^{[(l-j)n]}(i, z)\right\}=0.
\]
Clearly, in terms of KP wave function, we can rewrite the latter relation as
\[
J^{(n, p, l)}(i, z)
=\frac{z^l\psi_{i+p}-z^l\psi_{i+ln}-\sum_{j=1}^lz^{l-j}q_j^{(n, ln-p )}(i+p)\psi_{i+(l-j)n}}{\psi_i}=0.
\]
Thus in terms of the KP wave function the restriction of $n$th discrete KP hierarchy on ${\mathcal M}_{n, p, l}$ is given by the linear equation
\begin{equation}
z^l\psi_{i+ln}+\sum_{j=1}^lz^{l-j}q_j^{(n, ln-p )}(i+p)\psi_{i+(l-j)n}=z^l\psi_{i+p}.
\label{auxiliary1}
\end{equation}
Second linear equation which we should have in mind is (\ref{evolution}).

What we obtain is that on ${\mathcal M}_{n, p, l}$ the KP wave function satisfies a linear equation $\mathcal{T}_1(\psi_i)=\psi_i$ with the $l$-reduced operator $\mathcal{T}_1$ in the case $ln-p\leq -1$ while if $ln-p\geq 1$ then it satisfies the equation $\mathcal{S}_1(\psi_i)=\psi_i$ with $l$-reduced operator $\mathcal{S}_1$. So, let
\[
h=\left\{
\begin{array}{l}
p-ln\;\;\;\mbox{if}\;\;\; ln-p\leq -1; \\
ln-p\;\;\;\mbox{if}\;\;\; ln-p\geq 1.
\end{array}
\right.
\]
Let us denote
\[
q_j^{(n, ln-p )}(i)=T^j(i-h-(j-1)n),\;\;\; j=1,\ldots, l
\]
in the first case and
\[
q_j^{(n, ln-p )}(i)=(-1)^jS^j(i-(j-1)n),\;\;\; j=1,\ldots, l
\]
in the second case, respectively. Substituting the latter into (\ref{auxiliary1}) gives (\ref{l1}) and  (\ref{l2}). We observe that in both cases
\begin{equation}
q_j^{(n, -sh)}(i)=T^j_{s}(i-sh-(j-1)n)
\label{qjns1}
\end{equation}
and
\begin{equation}
q_j^{(n, sh)}(i)=(-1)^jS^j_{s}(i-(j-1)n).
\label{qjns}
\end{equation}

Let us turn now to equation (\ref{evolution}) defining an infinite number of flows of the $n$th discrete KP hierarchy and consider only the flows $\partial_{sh}$. With (\ref{qjns}) we obtain
\begin{eqnarray}
\partial_{sh}\psi_i&=&z^{sh}\psi_{i+shn}+\sum_{j=1}^{sh}z^{sh-j}q_j^{(n, shn)}(i)\psi_{i+(sh-j)n} \nonumber \\
&=&z^{sh}\left\{\psi_{i+shn}+\sum_{j=1}^{sh}(-1)^jz^{-j}S^j_{sn}(i-(j-1))\psi_{i+(sh-j)n}\right\} \nonumber \\
&=&z^{sh}\left(\mathcal{S}_{sn}\right)_{+, sh}(\psi_i).
\end{eqnarray}

Let us summarize what we have said above. Restricting the $n$th discrete KP hierarchy to $\mathcal{M}_{n, p, l}$ or more exactly its sub-hierarchy
$\{\partial_{sh} : s\geq 1\}$, where $h$ is defined by the triple $(n, p, l)$ as above, leads to $l$-reduced lattice systems (\ref{evolution111}) or
(\ref{evolution2}) in dependence of the sign of $ln-p$. In the following two subsections we show how   discrete
equations (\ref{disc1}), (\ref{disc2}), (\ref{disc3}) and (\ref{disc4}) appear in this framework and therefore  we clarify their meaning.


\subsection{Weaker invariance conditions}
We have shown above that some integrable lattices can be obtained by imposig some suitable conditions compatible with the flows of the $n$th discrete KP hierarchy on $\mathcal{M}_{n, p, l}$ and presented some examples known in the literature. In this section, we are going to show that there exist weaker invariance conditions. To this aim, we observe that the relations
\begin{eqnarray}
\partial_sQ_k^{(n, p, l)}(i)&=&
Q_{s+k}^{(n, p, l)}(i+sn)+\sum_{j=1}^sq_j^{(n, sn)}(i+p)Q_{s+k-j}^{(n, p, l)}(i+(s-j)n) \nonumber \\
&&-Q_{s+k}^{(n, p, l)}(i)-\sum_{j=1}^sq_j^{(n, sn)}(i-(s+k-j)n)Q_{s+k-j}^{(n, p, l)}(i) \label{dts1}
\end{eqnarray}
are valid by virtue of the $n$th discrete KP hierarchy. Let $\{I_k^{(n, p, l)} : k\geq1\}$ be some infinite set of quasi-homogeneous polynomials related with $\{Q_k^{(n, p, l)} : k\geq1\}$ by relations
\begin{equation}
Q_k^{(n, p, l)}=I_k^{(n, p, l)}+\sum_{j=1}^{k-1}\xi_{k-1, j}I_{k-j}^{(n, p, l)}
\label{QI}
\end{equation}
with some unknown coefficients $\xi_{k, s}$ to be defined . Consider equation (\ref{dts1}) on $Q_1^{(n, p, l)}=I_1^{(n, p, l)}$, that is,
\begin{eqnarray}
\partial_sQ_1^{(n, p, l)}(i)&=&
Q_{s+1}^{(n, p, l)}(i+sn)+\sum_{j=1}^sq_j^{(n, sn)}(i+p)Q_{s+1-j}^{(n, p, l)}(i+(s-j)n) \nonumber \\
&&-Q_{s+1}^{(n, p, l)}(i)-\sum_{j=1}^sq_j^{(n, sn)}(i-(s+1-j)n)Q_{s+1-j}^{(n, p, l)}(i). \label{dts2}
\end{eqnarray}
Substituting (\ref{QI}) into the right-hand side of (\ref{dts2}) gives
\begin{eqnarray}
\fl
\partial_sI_1^{(n, p, l)}(i)&=&I_{s+1}^{(n, p, l)}(i+sn)+\sum_{j=1}^s\xi_{s, j}(i+sn)I_{s-j+1}^{(n, p, l)}(i+sn)+q_1^{(n, sn)}(i+p) \nonumber \\
\fl
&&\times\left\{I_s^{(n, p, l)}(i+(s-1)n)+\sum_{j=1}^{s-1}\xi_{s-1, j}(i+(s-1)n)I_{s-j}^{(n, p, l)}(i+(s-1)n)\right\}+\cdots \nonumber \\
\fl
&&+q_s^{(n, sn)}(i+p)I_1^{(n, p, l)}(i)-I_{s+1}^{(n, p, l)}(i)-\sum_{j=1}^s\xi_{s, j}(i)I_{s-j+1}^{(n, p, l)}(i)-q_1^{(n, sn)}(i-sn) \nonumber \\
\fl
&&\times\left\{I_s^{(n, p, l)}(i)+\sum_{j=1}^{s-1}\xi_{s-1, j}(i)I_{s-j}^{(n, p, l)}(i)\right\}-\cdots-q_s^{(n, sn)}(i-n)I_1^{(n, p, l)}(i). \nonumber
\end{eqnarray}
Impose now the periodicity condition $I_k(i+n)=I_k(i)$ for all $k\geq 1$. With this condition
\begin{eqnarray}
\fl
\partial_sI_1^{(n, p, l)}(i)&=&\left\{\xi_{s, 1}(i+sn)+q_1^{(n, sn)}(i+p)-\xi_{s, 1}(i)-q_1^{(n, sn)}(i-sn)\right\}I_s^{(n, p, l)}(i) \nonumber \\
\fl
&&+\left\{\xi_{s, 2}(i+sn)+q_1^{(n, sn)}(i+p)\xi_{s-1, 1}(i+(s-1)n)+q_2^{(n, sn)}(i+p) \right. \nonumber \\
\fl
&&\left. -\xi_{s, 2}(i)-q_1^{(n, sn)}(i-sn)\xi_{s-1, 1}(i)-q_2^{(n, sn)}(i-(s-1)n)\right\}I_{s-1}^{(n, p, l)}(i)+\cdots \nonumber \\
\fl
&&+\left\{\xi_{s, s}(i+sn)+\sum_{j=1}^{s-1}q_j^{(n, sn)}(i+p)\xi_{s-j, s-j}(i+(s-j)n)+q_s^{(n, sn)}(i+p)  \right. \nonumber \\
\fl
&&\left. -\xi_{s, s}(i)-\sum_{j=1}^{s-1}q_j^{(n, sn)}(i-(s-j+1)n)\xi_{s-j, s-j}(i)-q_s^{(n, sn)}(i-n)\right\}I_1^{(n, p, l)}(i). \nonumber
\end{eqnarray}
Now we observe that if $\xi_{s, k}(i)=q_k^{(n, -p-(s-k+1)n)}(i+p)$, then the coefficients at $I_j^{(n, p, l)}$ in the last relation become identically zeros. So, let
\begin{equation}
Q_k^{(n, p, l)}(i)=I_k^{(n, p, l)}(i)+\sum_{j=1}^{k-1}q_j^{(n, -p-(k-j)n)}(i+p)I_{k-j}^{(n, p, l)}(i).
\label{Ik1}
\end{equation}
One can check that resolving these relations in favor of $I_k^{(n, p, l)}$ yields
\begin{equation}
I_k^{(n, p, l)}(i)=Q_k^{(n, p, l)}(i)+\sum_{j=1}^{k-1}a_j^{[-p-(k-j)n]}(i+p)Q_{k-j}^{(n, p, l)}(i).
\label{Ik}
\end{equation}
Taking into account (\ref{QJ}), we derive from (\ref{Ik})
\begin{equation}
I_k^{(n, p, l)}(i)=J_k^{(n, p, l)}(i)+\sum_{j=1}^{k-1}a_j^{[-p]}(i+p)J_{k-j}^{(n, p, l)}(i)
\label{IJ}
\end{equation}
and
\[
J_k^{(n, p, l)}=I_k^{(n, p, l)}+\sum_{j=1}^{k-1}a_j^{[p]}(i)I_{k-j}^{(n, p, l)}.
\]
In terms of the Laurent series relations, (\ref{IJ}) takes the form
\begin{eqnarray}
I^{(n, p, l)}(i, z)&=&z^pa^{[-p]}(i+p, z)J^{(n, p, l)}(i, z) \nonumber \\
&=&\frac{\psi_i}{\psi_{i+p}}J^{(n, p, l)}(i, z) \nonumber \\
&=&\frac{z^l\psi_{i+p}-z^l\psi_{i+ln}-\sum_{j=1}^lz^{l-j}q_j^{(n, ln-p )}(i+p)\psi_{i+(l-j)n}}{\psi_{i+p}}. \nonumber
\end{eqnarray}

Therefore, we have proved the following. If all $I_k^{(n, p, l)}$ defined by (\ref{Ik}) satisfy  the periodicity condition
\begin{equation}
I_k^{(n, p, l)}(i+n)=I_k^{(n, p, l)}(i)
\label{periodicity}
\end{equation}
then $I_1^{(n, p, l)}$ do not depend of all  evolution parameters $t_s$. Using similar reasoning, step-by-step, we can show that under this condition  in fact all $I_k^{(n, p, l)}$ for $k\geq 1$ are constants.
It is evident that by virtue of (\ref{Ik1}), the equations $I_k^{(n, p, l)}\equiv 0$ define the invariant submanifold ${\mathcal M}_{n, p, l}$ while  the periodicity condition  (\ref{periodicity}) is weaker than the latter condition. We observe that the coefficients of equations (\ref{dts1}) considered as quasi-linear equations with respect to $Q_k^{(n, p, l)}$ do not depend on $l$. This means that periodicity conditions  $\tilde{I}_k^{(n, p, l)}(i+n)=\tilde{I}_k^{(n, p, l)}(i)$ where
\begin{equation}
\tilde{I}_k^{(n, p, l)}\equiv I_k^{(n, p, l)}+\sum_{j=1}^lh_jI_k^{(n, p, l-j)}
\label{tildeI}
\end{equation}
with arbitrary constants $h_j$ also select some invariant submanifold. Let $d$ be some divisor of the number $n$, then, it is obvious that the periodicity conditions
\begin{equation}
\tilde{I}_k^{(n, p, l)}(i+d)=\tilde{I}_k^{(n, p, l)}(i)
\label{invariant1}
\end{equation}
define some invariant submanifold ${\mathcal N}_{n, p, l}^d$ of the $n$th discrete KP  hierarchy. Let us formulate all the above-mentioned in the following theorem.

\begin{thm} \label{thm:general}
Infinite number of the periodicity conditions (\ref{invariant1}), where polynomials $\tilde{I}_k^{(n, p, l)}$ are defined by (\ref{Ik}) and (\ref{tildeI}) and $d$ is some divisor of $n$, are compatible with the flows of the $n$th discrete KP hierarchy.
\end{thm}


\subsection{Discrete equations (\ref{disc1}), (\ref{disc2}), (\ref{disc3}) and (\ref{disc4}) as reductions of evolution differential-difference equations}
Here,  we use the quite general theorem \ref{thm:general} to construct classes of constraints compatible with integrable hierarchies corresponding to different submanifolds $\mathcal{M}_{n, p, l}$ .
Let the triple of integers $(n, \tilde{p}, k)$ be such that $kn-\tilde{p}=sh$ for some arbitrary $s\geq 0$, then by virtue of (\ref{qjns})
\begin{eqnarray}
\fl
I^{(n, \tilde{p}, k)}(i, z)&=&\frac{z^k\psi_{i+\tilde{p}}-z^k\psi_{i+kn}-\sum_{j=1}^kz^{k-j}q_j^{(n, kn-\tilde{p})}(i+\tilde{p})\psi_{i+(k-j)n}}{\psi_{i+\tilde{p}}} \nonumber \\
\fl
&=&\frac{z^k\psi_{i+\tilde{p}}-z^k\psi_{i+kn}-\sum_{j=1}^k(-1)^jz^{k-j}S^j_s(i+\tilde{p}-(j-1)n)\psi_{i+(k-j)n}}{\psi_{i+\tilde{p}}}. \nonumber
\end{eqnarray}
Replacing $k\mapsto k-qh$, $s\mapsto s-qn$ and therefore $\tilde{p}\mapsto \tilde{p}$, we obtain
\begin{eqnarray}
\fl
I^{(n, \tilde{p}, k-qh)}(i, z)&=&\left\{z^{k-qh}\psi_{i+\tilde{p}}-z^{k-qh}\psi_{i+(k-qh)n}             \phantom{\sum_{j=1}^{k-qh}}\right. \nonumber \\
\fl
&&\left.-\sum_{j=1}^{k-qh}(-1)^jz^{k-qh-j}S^j_{s-qn}(i+\tilde{p}-(j-1)n)\psi_{i+(k-qh-j)n}\right\}/\psi_{i+\tilde{p}} \nonumber
\end{eqnarray}
and consequently
\begin{eqnarray}
\fl
\tilde{I}^{(n, \tilde{p}, k)}(i, z)&\equiv& I^{(n, \tilde{p}, k)}(i, z)+\sum_{q=1}^{\kappa}H_qI^{(n, \tilde{p}, k-qh)}(i, z) \nonumber \\
\fl
&=&\left\{\left(z^k+\sum_{q=1}^{\kappa}H_qz^{k-qh}\right)\psi_{i+\tilde{p}}\right. \nonumber \\
\fl
&&\left.-z^k\psi_{i+kn}-\sum_{j=1}^k(-1)^jz^{k-j}\tilde{S}^j_s(i+\tilde{p}-(j-1)n)\psi_{i+(k-j)n}\right\}/\psi_{i+\tilde{p}}. \label{const1}
\end{eqnarray}
Impose the condition
\[
\tilde{I}^{(n, \tilde{p}, k)}(i, z)=\sum_{j\geq 1}\tilde{I}_jz^{-j},
\]
where $\tilde{I}_j$'s are some constants. According to theorem \ref{thm:general} this condition is invariant with respect to the flows of the $n$th discrete KP hierarchy. One sees that (\ref{const1}) becomes linear problem (\ref{lin222}), with
\[
w=z^k+\sum_{q=1}^{\kappa}H_qz^{k-qh}-\sum_{j\geq 1}\tilde{I}_jz^{-j}.
\]

By analogy, let the triple of integers $(n, \bar{p}, k)$ be such that $kn-\bar{p}=-sh$ for some arbitrary $s\geq 0$. Then by virtue of (\ref{qjns1})
\begin{eqnarray}
\fl
I^{(n, \bar{p}, k)}(i, z)&=&\frac{z^k\psi_{i+\bar{p}}-z^k\psi_{i+kn}-\sum_{j=1}^kz^{k-j}q_j^{(n, kn-\bar{p})}(i+\bar{p})\psi_{i+(k-j)n}}{\psi_{i+\tilde{p}}} \nonumber \\
\fl
&=&\frac{z^k\psi_{i+\bar{p}}-z^k\psi_{i+kn}-\sum_{j=1}^kz^{k-j}T^j_s(i+(k-j+1)n)\psi_{i+(k-j)n}}{\psi_{i+\bar{p}}}. \nonumber
\end{eqnarray}
Imposing invariant condition
\[
\tilde{I}^{(n, \bar{p}, k)}(i, z)\equiv I^{(n, \bar{p}, k)}(i, z)+\sum_{q=1}^{\kappa}c_qI^{(n, \bar{p}, k-qh)}(i, z)=\sum_{j\geq 1}\tilde{I}_jz^{-j}
\]
leads to linear problem (\ref{lin333}).


\subsection{Examples}

\subsubsection{One-field difference systems}

In the case $\mathcal{M}_{n, n+1, 1}$ we have $ln-p=-1$ and therefore $h=p-ln=1$. In this situation the $n$th discrete KP hierarchy is reduced to the Itoh-Narita-Bogoyavlenskii lattice given by (\ref{INB}). Two classes of compatible constraints are deduced as the one-field reduction of discrete equations (\ref{disc1}) and (\ref{disc3}), that is,
\[
\tilde{T}^k_s(i)T(i+s+kn)=\tilde{T}^k_s(i+n+1)T(i)
\]
and
\[
\tilde{S}^k_s(i)T(i+kn)=\tilde{S}^k_s(i+n+1)T(i+s),
\]
respectively, with
\[
\tilde{T}^k_s(i)=T^k_s(i)+\sum_{j=1}^{k}c_jT^{k-j}_{s+jn}(i)
\]
and
\[
\tilde{S}^k_s(i)=S^k_s(i)+\sum_{j=1}^{k}(-1)^{j}H_jS^{k-j}_{s-jn}(i+jn).
\]
Remember that we suppose here that $s-jn\leq 1$. A more general class of one-field lattices is given by
\begin{equation}
\tilde{T}^k_s(i)T(i+sh+kn)=\tilde{T}^k_s(i+n+h)T(i)
\label{T111}
\end{equation}
and
\begin{equation}
\tilde{S}^k_s(i)T(i+kn)=\tilde{S}^k_s(i+n+h)T(i+sh),
\label{S111}
\end{equation}
where $h$ and $n$ are supposed to be co-prime positive integers.  Both the equations (\ref{S111}) and (\ref{T111}) are of the form of the autonomous difference equation (\ref{ODE})
with the order $N=sh+kn$ whose right-hand side is a rational function of its arguments and corresponding parameters. Let us observe that making use of the identities for polynomials $\tilde{T}^k_s$ and $\tilde{S}^k_s$ we can rewrite discrete equations (\ref{S111}) and (\ref{T111}) in the following equivalent form:
\begin{equation}
\tilde{T}^k_{s-1}(i+h)T(i+sh+kn)=\tilde{T}^k_{s-1}(i+n+h)T(i)
\label{T222}
\end{equation}
and
\begin{equation}
\tilde{S}^k_{s+1}(i)T(i+kn)=\tilde{S}^k_{s+1}(i+n)T(i+sh).
\label{S222}
\end{equation}
It is evident from this that in the case $n=0$ and $h=1$ which corresponds to symmetric multi-variate polynomials  $T^k_s\{y_j\}$ and $S^k_s\{y_j\}$, equations (\ref{S222}) and (\ref{T222}) are specified only as the periodicity equation $T(i+s)=T(i)$. 


\subsubsection{The case $h=n=1$}
The simplest case, when we can obtain nontrivial discrete nonlinear equations different from the periodicity one is $h=n=1$. In this case (\ref{S222}) and (\ref{T222}) looks as
\begin{equation}
\tilde{T}^k_{s-1}(i+1)T(i+s+k)=\tilde{T}^k_{s-1}(i+2)T(i)
\label{T333}
\end{equation}
and
\begin{equation}
\tilde{S}^k_{s+1}(i)T(i+k)=\tilde{S}^k_{s+1}(i+1)T(i+s).
\label{S333}
\end{equation}
What we know is that these equations play the role of compatible constraints for the Volterra lattice hierarchy defining invariant submanifolds in solution space of the Volterra lattice. We claim but do not prove here
that for an arbitrary fixed pair $(k, s)$ such that $s\geq k+1$,  (\ref{S333})  and (\ref{T333}) represents in fact the same dynamical system. The point is that one can transform (\ref{S333}) into (\ref{T333}) and back by interchanging $\{c_1,\ldots, c_k\}$ and $\{H_1,\ldots, H_k\}$ (cf. \cite{Roberts}). We will prove  this  fact in  a subsequent publication.

For spectral curves, our calculations using Lax operators defined by (\ref{lin333}) and (\ref{l1}), in the case $h=n=l=1$, yield the following. If $s+k=2g+1$, where $g\geq k$ we obtain
\begin{equation}
H_0w^2+z^{g+1}R_g(z)w-z^{2g+1}\tilde{R}_k(z)=0,
\label{curve1}
\end{equation}
while in the case $s+k=2g+2$ , where $g\geq k$, we get 
\begin{equation}
H_0w^2-z^{g+1}R_{g+1}(z)w+z^{2g+2}\tilde{R}_k(z)=0,
\label{curve2}
\end{equation}
with $R_g(z)=z^g+\sum_{j=1}^gH_jz^{g-j}$ and $\tilde{R}_k(z)=z^k+\sum_{j=1}^k\tilde{H}_jz^{k-j}$,
where $\tilde{H}_j=H_j+\sum_{q=1}^{j-1}c_qH_{j-q}+c_j$. We remark that $H_j$'s are the first integrals integrals of equation (\ref{T333}), while $c_j$'s are the parameters entered in these equations. 
In particular, 
\begin{equation}
H_0=\frac{\prod_{j=1}^{s+k}T_{i+j-1}}{\tilde{T}^k_{s-1}(i+1)}.
\label{H0}
\end{equation}
In both cases we have an expansion
\[
w=z^k+\sum_{j=1}^kc_jz^{k-j}-\sum_{j\geq 1}\tilde{I}_jz^{-j},
\]
where the coefficients $\tilde{I}_j$ are some polynomials of variables $H_j$ and $c_j$. 

Let us remark that in both cases we have a hyperelliptic spectral curve of genius $g$. For (\ref{curve1}), with the help of birational transformation
\[
z=\frac{1}{Z},\;\;
w=\frac{W-r_g(Z)}{2H_0Z^{2g+1}},\;\;
W=\frac{2H_0w+z^{g+1}R_g(z)}{2z^{2g+1}},
\]
where
\[
r_g(Z)\equiv Z^gR_g\left(\frac{1}{Z}\right),\;\;
\tilde{r}_k(Z)\equiv Z^k\tilde{R}_k\left(\frac{1}{Z}\right),
\]
we get
\[
W^2=P_{2g+1}(Z)\equiv r_g^2(Z)+4H_0Z^{2g-k+1}\tilde{r}_k(Z),
\]
while for (\ref{curve2}), we use 
\[
z=Z,\;\; 
w=Z^{g+1}\frac{W+R_{g+1}(Z)}{2H_0},\;\; 
W=\frac{2H_0w-z^{g+1}R_{g+1}(z)}{z^{g+1}}
\]
to obtain
\[
W^2=P_{2g+2}(Z)\equiv R_{g+1}^2(Z)-4H_0\tilde{R}_k(Z).
\]

Making use of the transformation 
\[
w\mapsto \frac{z^{g+1}}{H_0}\left(z^{g-k}w-R_g(z)\right)
\]
for (\ref{curve1}) gives
\begin{equation}
z^{2g-2k+1}w^2-z^{g-k+1}R_g(z)w-H_0\tilde{R}_k(z)=0
\label{curve11}
\end{equation}
while the transformation
\[
w\mapsto \frac{z^{g+1}}{H_0}\left(R_{g+1}(z)-z^{g-k+1}w\right)
\]
converts (\ref{curve2}) to the following form:
\begin{equation}
z^{2g-2k+2}w^2-z^{g-k+1}R_{g+1}(z)w+H_0\tilde{R}_k(z)=0.
\label{curve22}
\end{equation}
The reason why we have shown these transformations is that (\ref{curve11}) and (\ref{curve22}) are nothing but the spectral curves for Lax operators defined by  (\ref{l1}) and (\ref{lin222}), in the case $h=n=l=1$. Let us remark that in this case $\{H_1,\ldots, H_k\}$ is the set of the parameters entered  in equation (\ref{S333}), while other $H_j$'s and all $c_j$'s are the first integrals. We will describe in full detail all these integrals in a subsequent publication. 


\subsection{Bibliographical remarks}

Let us remark that (\ref{S333}) has obvious integral\footnote{Here we use the same symbol $H_0$ as in (\ref{H0}) because as we have already said equations (\ref{S333})  and (\ref{T333}) for fixed pair $(k, s)$ represent the same dynamical system. Therefore the integral $H_0$ has two different, at first glance, forms (\ref{H0}) and (\ref{Quispel}).}
\begin{equation}
H_0=(-1)^k\tilde{S}^k_{s+1}(i)\prod_{j=k}^{s-1}T(i+j).
\label{Quispel}
\end{equation}
In particular case $k=1$ one has
\[
H_0=-\left\{S^1_{s+1}(i)-H_1\right\}\prod_{j=1}^{s-1}T(i+j),
\]
where $S^1_{s+1}(i)=\sum_{j=0}^{s}T(i+j)$. The latter equation restricted by the condition  $H_1=0$ together with its integrals can be found  in \cite{Demskoi} (eq. (4)). In particular, in that paper plausible formula was found yielding degrees of mapping iterates suggesting that this equation has zero algebraic entropy.

\section{Discussion}
Studies carried out in \cite{Demskoi}, \cite{Kamp},  \cite{Svinin1}, \cite{Svinin6}, \cite{Tran1} and in this paper suggest that the polynomials $T^k_s$ and $S^k_s$ corresponding to different pairs of co-prime integers $n$ and $h$ might be quite universal and convenient combinatorial object and therefore we hope that they deserve attention to investigate them.

Unfortunately, in this paper we have not considered in full detail the important question of the structure of the first integrals for the obtained ordinary difference equations and corresponding spectral curves  $P(w, z)=0$. We  hope to fill this gap in the following publications to investigate one-field equations (\ref{S111}) and (\ref{T111}) and, in particular, equations (\ref{S333}) and (\ref{T333}). Also, we would like to study the relationship of the explicit form of integrable hierarchies (\ref{evolution111}) and (\ref{evolution2}) with their Hamiltonian structures. 

\section*{Acknowledgments}
I wish to thank the referees for carefully reading the manuscript and for remarks which enabled the presentation of the paper to be improved. This work was partially supported by grant NSh-5007.2014.9.

\end{document}